\documentclass[reprint,aps,prl,twocolumn,preprintnumbers,amsmath,amssymb,floatfix]{revtex4}

\usepackage{epsfig}
\usepackage[colorlinks=true,linkcolor=blue]{hyperref} 
\usepackage{xcolor}
\usepackage{ulem}
\usepackage{dcolumn}
\usepackage{physics}
\usepackage{multirow}

\usepackage{pifont}% http://ctan.org/pkg/pifont

\DeclareGraphicsExtensions{.pdf}

\begin{document}

\title{Fragmentation Functions of Charged Hadrons at Next-to-Next-to-Leading Order and Constraints on the Proton Parton Distribution Functions}

\author{Jun~Gao$^{1}$,~XiaoMin~Shen$^{2,1}$,~Hongxi Xing$^{3,4,5}$,~Yuxiang Zhao$^{2,5,6,7}$,~Bin~Zhou$^{1}$}

\affiliation{
    $^1$State Key Laboratory of Dark Matter Physics, Shanghai Key Laboratory for Particle Physics and Cosmology, Key Laboratory for Particle Astrophysics and Cosmology (MOE), School of Physics and Astronomy, Shanghai Jiao Tong University, Shanghai 200240, China\\
    $^2${Institute of Modern Physics, Chinese Academy of Sciences, Lanzhou, Gansu 730000, China}\\
    $^3${State Key Laboratory of Nuclear Physics and Technology, Institute of Quantum Matter, South China Normal University, Guangzhou 510006, China}\\
    $^4${Guangdong Basic Research Center of Excellence for Structure and Fundamental Interactions of Matter, Guangdong Provincial Key Laboratory of Nuclear Science, Guangzhou 510006, China}\\
    $^5${Southern Center for Nuclear-Science Theory (SCNT), Institute of Modern Physics, Chinese Academy of Sciences, Huizhou 516000, China}\\
    $^6${University of Chinese Academy of Sciences, Beijing 100049, China}\\
    $^7${Key Laboratory of Quark and Lepton Physics (MOE) and Institute of Particle Physics, Central China Normal University, Wuhan 430079, China}
 }

\email{\\ {jung49@sjtu.edu.cn} \\
{xiaominshen@impcas.ac.cn}\\
{hxing@m.scnu.edu.cn}\\
{yxzhao@impcas.ac.cn}\\
{zb0429@sjtu.edu.cn}\\}

\begin{abstract}
We present the first global analysis of fragmentation functions (FFs) for light charged hadrons ($\pi^{\pm}$, $K^{\pm}$) at full next-to-next-to-leading order in Quantum Chromodynamics (QCD), incorporating world data from both single-inclusive electron-positron annihilation and semi-inclusive deep-inelastic scattering.
The collinear factorization has been tested with low-momentum-transfer data and has demonstrated success at high hadron momenta.
Additionally, we study the impact of current global data on hadron production to the parton distribution functions (PDFs), and find they favor a reduced asymmetry in the strange (anti-) quark PDFs, as compared to the asymmetry predicted by state-of-the-art PDFs derived from inclusive data.
\end{abstract}
 \maketitle

\pagebreak
\newpage

\noindent{\it Introduction.--}
Fragmentation functions (FFs) characterize the probability density of a quark or gluon transitioning into color-neutral hadrons, expressed in terms of the light-cone momentum fraction. This concept was introduced and explained in detail by Field and Feynman in Refs.~\cite{Berman:1971xz,Field:1977fa,Feynman:1978dt}. 
In addition to their fundamental role in understanding color confinement in QCD, FFs are crucial for probing the internal structure of nucleons. This becomes especially important in the upcoming precision era of high-energy nuclear physics, driven by the development of electron-ion colliders (EIC)~\cite{Accardi:2012qut,Anderle:2021wcy}.
For semi-inclusive hadron production in deep-inelastic scatterings (SIDIS), a key process at the upcoming EICs, the cross sections can be factorized as hard scattering matrix elements convoluted with non-perturbative quantities, including both parton distribution functions (PDFs) and FFs~\cite{Collins:1989gx}.
State-of-the-art results for PDFs have been extracted through global data analyses with next-to-next-to-leading order (NNLO) accuracy~\cite{Alekhin:2017kpj,Hou:2019efy,Bailey:2020ooq,NNPDF:2021njg,ATLAS:2021vod} and approximate next-to-next-to-next-to-leading order accuracy~\cite{McGowan:2022nag,NNPDF:2024nan} in QCD.
In contrast, although significant efforts have been devoted to determining FFs at NNLO using data solely from single-inclusive electron-position annihilation (SIA)~\cite{Bertone:2017tyb,Soleymaninia:2018uiv}, the global analysis of FFs remains at an approximate NNLO accuracy considering both SIA and SIDIS data~\cite{Borsa:2022vvp,AbdulKhalek:2022laj}.
In this letter, we present the first determination of FFs for charged pions and kaons at full NNLO accuracy in QCD, based on a global analysis of data from SIA and SIDIS.
Building on advancements in theoretical precision, we are able to assess the consistency of identified charged hadron production between
$e^+e^-$ collisions and SIDIS measurements, enabling a robust test of QCD factorization at low energy scales by incorporating recent SIA data from BESIII~\cite{BESIII:2025aaa}.
We find good agreements between our theoretical predictions and the data sets included in our fit, considering residual theoretical uncertainties.
{Furthermore, new developments have been made to study the impact of global data on hadron production to the PDFs at NNLO accuracy in QCD for the first time.}
{Leveraging current state-of-the-art PDFs as baselines, we find a preference for reduced asymmetry between strange quark and anti-quark distributions.}
\noindent{\it Theoretical setup and data characteristics.--}
The FFs for $\pi^+$ and $K^+$ are parametrized at an initial scale of $Q_0=1.4$ GeV using the same functional form in terms of the momentum fraction $z$ as adopted in Refs.~\cite{Gao:2024nkz,Gao:2024dbv}. 
{We have assumed certain flavor symmetries at the initial scale among favored and unfavored quarks
which leads to a total of 54 free parameters in the analysis (see Supplemental Material~\cite{fPARA}).}
The FFs are evolved to higher scales using three-loop
time-like splitting kernels~\cite{Mitov:2006ic,Moch:2007tx,Almasy:2011eq,Chen:2020uvt}, implemented within a modified version of HOPPET~\cite{Salam:2008qg}, to ensure consistency with the NNLO analysis.
Heavy quark FFs are
non-zero but do not evolve until the mass thresholds are reached, specifically at $m_c\!=\!1.4~{\rm GeV}$ and $m_b=4.5~{\rm GeV}$ for charm and bottom quarks.
Their contributions to low-energy SIDIS are suppressed by the heavy-quark PDFs,
while those to SIA are open only at energies above the threshold for heavy meson pair production. 
{We note recent progress on matching conditions of FFs at NNLO at the heavy-quark thresholds~\cite{Biello:2024zti}.}
Theoretical calculations of differential cross sections are carried out at NNLO in QCD using the FMNLO program~\cite{Liu:2023fsq,Zhou:2024cyk},
which relies on perturbative coefficient functions for SIDIS calculated in \cite{Goyal:2023zdi,Bonino:2024qbh} and for SIA in~\cite{Rijken:1996vr,Rijken:1996npa,Rijken:1996ns,Mitov:2006wy,Soar:2009yh,Almasy:2011eq,Xu:2024rbt}. 
These calculations are accelerated through the use of interpolation grids and fast convolution algorithms.
For calculations involving initial hadrons, the CT18 NNLO PDFs with $\alpha_S(M_Z)=0.118$~\cite{Hou:2019efy} are used. 
We set the renormalization, factorization and fragmentation scales to be the same, with a nominal value of the momentum transfer $Q$ for both SIA and SIDIS.
Theoretical uncertainties are included in the covariance matrix of $\chi^2$ calculations, and are assumed to be fully correlated among data sets measured at similar energies.
These uncertainties are estimated by the half width of the envelope of theoretical predictions when varying all the QCD scales simultaneously by a factor of two. 
We consider data sets from SIA and SIDIS with kinematic cuts of $Q>2$~GeV and $z>0.01$,
with the hadron energy fraction defined as usual, i.e., 
$z \equiv 2E_h/Q$ for SIA, and
$z \equiv (P\cdot P_h)/(P\cdot q)$ for SIDIS. 
To further test the validity of QCD collinear factorization, we impose an additional requirement on the hadron energy, setting a lower bound of $E_{h,{\rm min}}$, which will be varied during the analysis.
The hadron energies are measured in the center-of-mass (Breit) frame for SIA (SIDIS).
For SIA at low energies, we assume that factorization applies to the light-cone momentum fraction of the hadron, consistent with the treatment in SIDIS~\cite{Field:1977fa,MoosaviNejad:2015lgp}. %However, this assumption involves corrections due to the finite mass of the hadron when translating to measured distributions in terms of hadron momentum or energy~\cite{MoosaviNejad:2015lgp}.
For SIDIS, we utilize measurements on production of identified charged hadrons from COMPASS with isoscalar target~\cite{COMPASS:2016xvm,COMPASS:2016crr}.
For SIA, we incorporate a comprehensive set of data from 
Belle, BaBar, TASSO and TPC~\cite{Belle:2013lfg,BaBar:2013yrg,TASSO:1988jma,TPCTwoGamma:1988yjh} below the $Z$-pole, from OPAL, ALEPH, DELPHI, and SLD at the $Z$-pole~\cite{OPAL:1994zan,ALEPH:1994cbg,DELPHI:1998cgx,SLD:2003ogn}, and from OPAL and DELPHI above the $Z$-pole~\cite{OPAL:2002isf,DELPHI:2000ahn}.
Importantly, we further include
SIA measurements of $\pi^{\pm}$ and $K^{\pm}$ production from BESIII~\cite{BESIII:2025aaa} with center-of-mass energies of 3.05, 3.5 and 3.671 GeV. 
These energies are much lower than the aforementioned datasets used in existing global analyses, providing a more thorough test of collinear factorization. 
The correlations of experimental uncertainties have been carefully accounted for, as they can significantly influence the resulting $\chi^2$ values.

\noindent{\it Results on FFs.--}
In order to test the impact of finite hadron masses and other potential power corrections on QCD collinear factorization, we perform a scan over the hadron energy cut $E_{h,{\rm min}}$ in a range of 0.5 to 1~GeV.
In particular, we separate all data sets into four groups, i.e., SIA data from BESIII, B-factories, SIA data measured at high energies ($\sim 20-200$~GeV, HE-SIA), and SIDIS data from COMPASS.
The number of data points and $\chi^2$ for individual data set as well as global data are summarized in Tab.~\ref{tab:chi2}. 
An overall good agreement between NNLO predictions of our fits and the experimental data is observed.
We notice a quick growth of $\chi^2/N_{\rm pt}$ for the global data when $E_{h,{\rm min}}<0.8$~GeV, as well as for individual data set, especially the COMPASS data.
{The trends on deterioration of $\chi^2$ are similar for fits with pion (kaon) data only, or for fits at NLO accuracy (see Supplemental Material~\cite{fSCAN}).}
This hints a boundary where deviations from leading twist collinear factorization begin to emerge due to finite momentum and mass effects. Consequently, we adopt a nominal choice of $E_{h,{\rm min}}=0.8$~GeV in our following analyses.
For this choice, the $\chi^2$ of the global data is about 
781.8 units for a total number of data points of 919.
The agreement is good with $\chi^2/N_{\rm pt}$ slightly below 1, and the maximum of effective Gaussian measure~\cite{Hou:2019efy} of all data groups is 2.32$\sigma$, arising from the COMPASS data. 
{In Tab.~\ref{tab:chi2} we also show quality of parallel fits at NLO and at approximate NNLO using the threshold results on SIDIS cross sections~\cite{Abele:2021nyo}.
Both fits lead to slightly worse $\chi^2$ of the global data.
Besides, the description of COMPASS data improves marginally when excluding the BESIII data from the fit. }

\begin{table*}[t]
  \
  
  \begin{tabular}{|c|c|c|c|c|c|c|c|c|c||c|c|c|}
    \hline
     \multirow{2}{*}{} & \multirow{2}{*}{${E_{h,{\rm min}}}$$[{\rm{GeV}}]$} &  \multicolumn{2}{c|}{BESIII} &  \multicolumn{2}{c|} {COMPASS} & \multicolumn{2}{c|} {B-factories} & \multicolumn{2}{c||} {HE-SIA} &  \multicolumn{3}{c|} {global} \\
    \cline{3-13}
    &  & $N_{\rm{pt}}$ &
    ${\chi^2}/{N_{\rm{pt}}}$ & $N_{\rm{pt}}$ &
    ${\chi^2}/{N_{\rm{pt}}}$ & $N_{\rm{pt}}$ &
    ${\chi^2}/{N_{\rm{pt}}}$ & $N_{\rm{pt}}$ & 
    ${\chi^2}/{N_{\rm{pt}}}$ & $N_{\rm{pt}}$ & $\chi^2$ &
    ${\chi^2}/{N_{\rm{pt}}}$\\
    \hline
    \multirow{6}{*}{NNLO} &
  0.5 & 242 & 1.26 & 358 & 1.65 & 233 & 1.06 & 426 & 1.19 & 1259 & 1650.2 &
  1.31\\
  &0.6 & 212 & 1.21 & 290 & 1.59 & 228 & 0.92 & 423 & 0.97 & 1153 & 1338.8 &
  1.16\\
  &0.7 & 182 & 1.11 & 214 & 1.47 & 223 & 0.61 & 413 & 0.84 & 1032 & 997.2 &
  0.97\\
  &0.8 & 152 & 0.98 & 142 & 1.30 & 218 & 0.53 & 407 & 0.82 & 919 & 781.8 &
  0.85\\
  &0.9 & 122 & 1.05 & 94 & 1.29 & 213 & 0.52 & 407 & 0.80 & 836 & 687.1 &
  0.82\\
  &1.0 & 98 & 1.14 & 54 & 0.97 & 209 & 0.49 & 403 & 0.80 & 764 & 587.2 & 0.77\\
  \hline
  {NLO} &0.8 & 152 & 1.03 & 142 & 1.26 & 218 & 0.54 & 407 & 0.85 & 919 & 801.6 & 0.87\\
  {NNLO(approx.)}&0.8 & 152 & 0.96 & 142 & 1.40 & 218 & 0.53 & 407 & 0.81 & 919 & 791.5 & 0.86 \\
  {NNLO(w/o BES)} & 0.8 & - & - & 142 & 1.23 & 218 & 0.52 & 407 & 0.81 & 767 & 620.2 & 0.81\\
    \hline
  \end{tabular}
  \caption{Quality of fit for different choices of lower cut on the hadron
  energy {at various accuracy, and of fit without BESIII data.}
  }\label{tab:chi2}
\end{table*}

\begin{figure}[t]
    \includegraphics[width=1.0\linewidth]{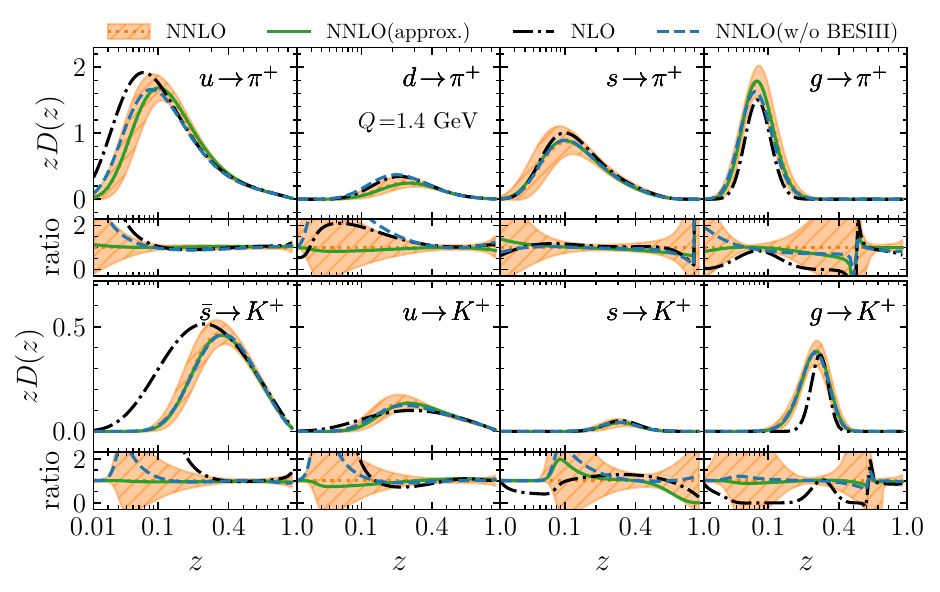}
    \caption{Fragmentation functions of $\pi^+$ and $K^+$ for diverse partons at the starting scale $Q_0=1.4~{\rm GeV}$.
    The dotted lines represent the nominal fit, while the uncertainties at $68\%$ C.L. are shown in colored band.
    {Other lines represent alternative fits as listed in Tab.~\ref{tab:chi2}.}
    }
    \label{fig:ffs}
\end{figure}

The extracted NNLO FFs at the initial scale for positively charged pion and kaon are shown in Fig.~\ref{fig:ffs} as functions of the momentum fraction $z$ for light quarks and gluon.
We plot the Hessian uncertainty bands together with the best-fit FFs, both in absolute values and by normalized to the best-fit FFs.
We use a tolerance of $\Delta\chi^2=2.32^2\approx 5.4$, namely square of the maximal effective Gaussian variables of all groups, in our estimation of uncertainties of the FFs with the Hessian method~\cite{Kovarik:2019xvh}.
The pion FFs from the favored quarks are well constrained at large-$z$ region where most of the data was measured, similarly for kaon FFs from the $\bar s$ and $u$ quarks.
For $z<0.1$, the FFs are mostly constrained by SIA data measured at or above the $Z$ boson mass, which can probe $z$ values as low as 0.02 after the kinematic selection.
At the initial scale, the FFs from unfavored quarks are nearly negligible, except for the strange quarks fragmenting into pions. Meanwhile, the FFs from gluons exhibit a pronounced peak and are indirectly constrained through scaling violations.
{We further compare our nominal NNLO FFs with those from alternative fits in Fig.~\ref{fig:ffs}.
It shows that FFs from the approximate NNLO fit are close to our nominal ones, while FFs from the NLO fit show a large discrepancy.
FFs from the fit without BESIII data reveal moderate pulls on $u/d$ quark and gluon to pions from BESIII measurements.
Comparisons of our FFs to previous determinations at approximate NNLO are also available (see Supplemental Material~\cite{fCOM}).}

\begin{figure}[t]
    \centering
    \includegraphics[width=1.0\linewidth]{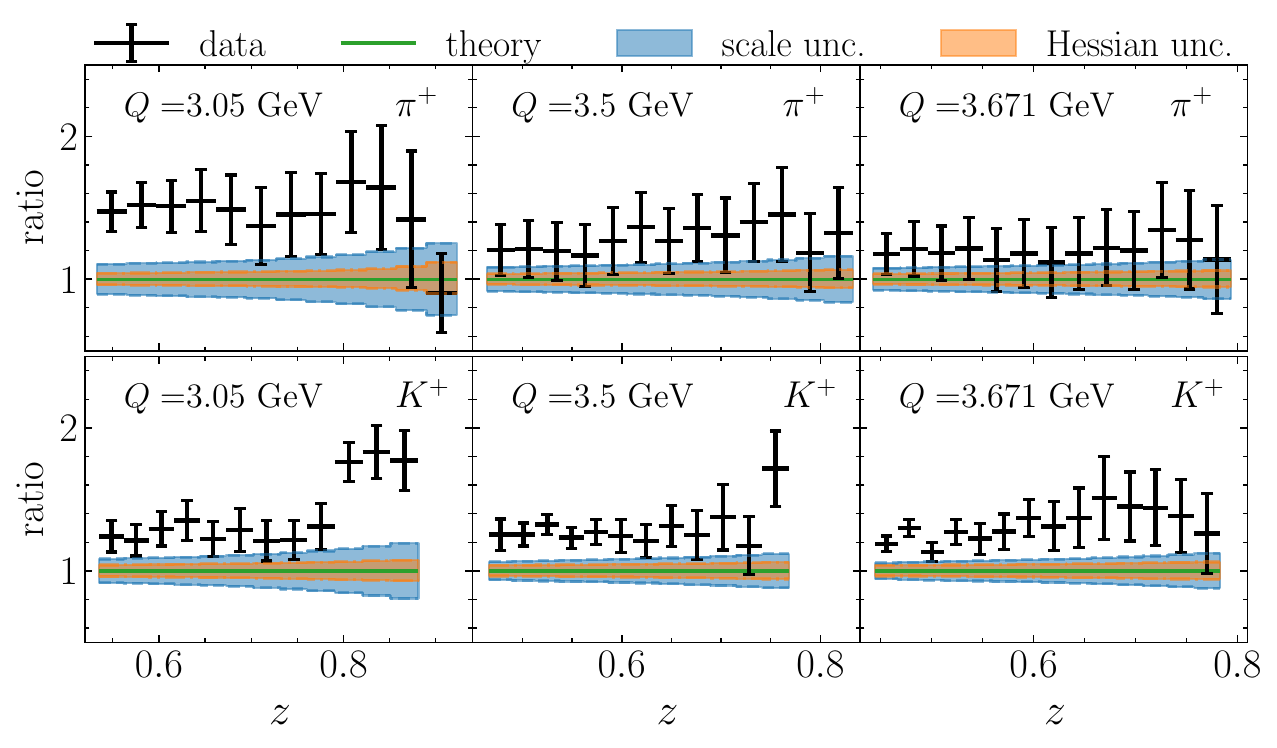}
    \caption{Comparison of theory and data
    for $\pi^+$ and $K^+$ measurements at BESIII.
    The results are normalized to theoretical predictions.
    $Q$ is the center-of-mass energy.
    }
    \label{fig:bes}
\end{figure}

\begin{figure}
    \centering
    \includegraphics[width=1.0\linewidth]{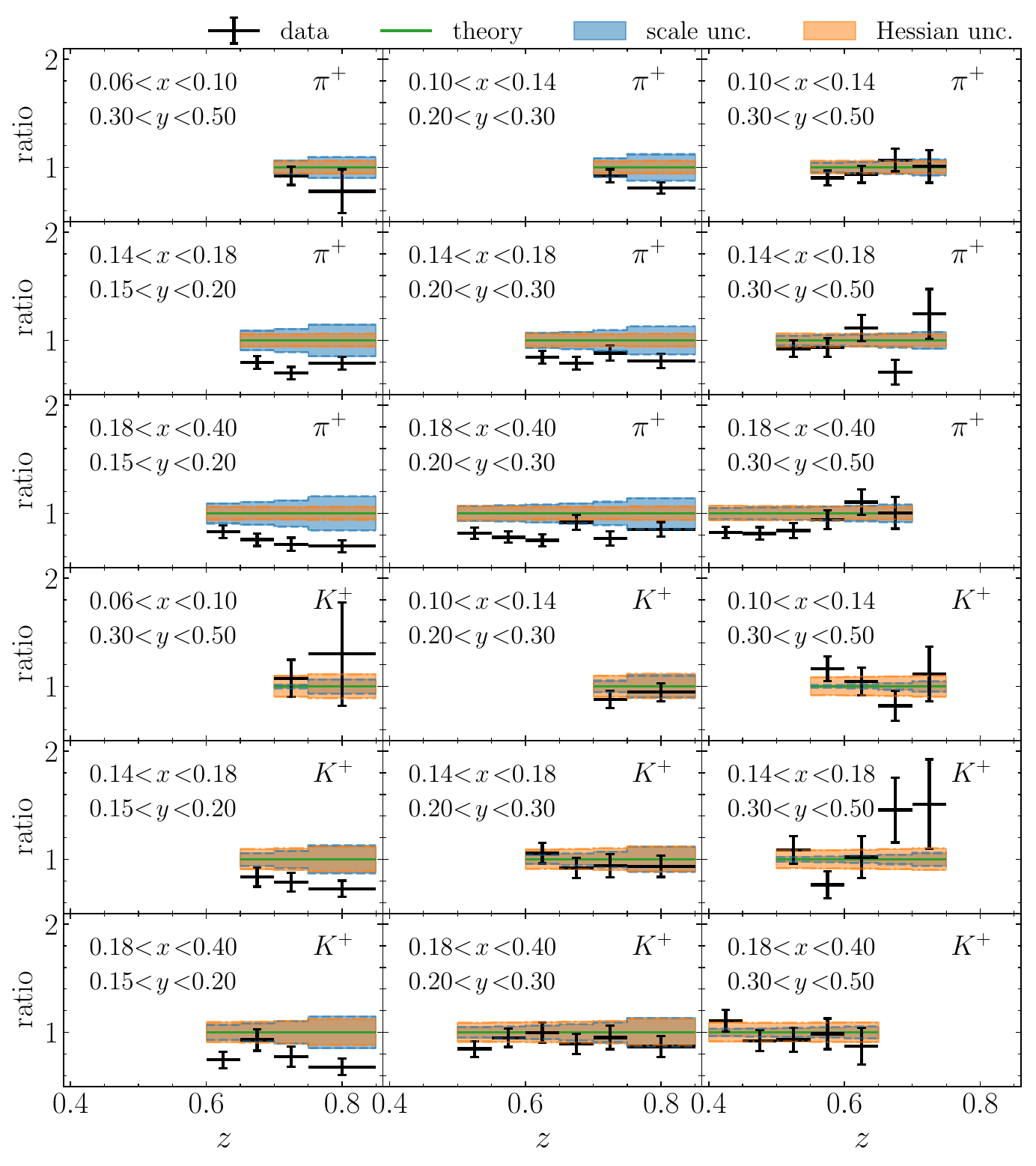}
    \caption{Similar to Fig.~\ref{fig:bes}, but for multiplicity measurements in SIDIS from the COMPASS Collaboration for different bins of Bjorken-$x$ and inelasticity $y$.
    }
    \label{fig:compass}
\end{figure}

Importantly, developments in both theory and experiment now enable us to test QCD factorization at low $Q$, around a few GeV. This is achieved by examining the consistency between SIA and SIDIS data through NNLO analysis.
Figs.~\ref{fig:bes} and~\ref{fig:compass} present detailed comparisons between our nominal NNLO predictions and the BESIII and COMPASS measurements of $\pi^+$ and $K^+$ multiplicities, shown as functions of hadron energy fraction $z$.
All results are normalized to the central values of theoretical predictions.
The error bars represent the total experimental uncertainties while the colored bands are the scale uncertainties and the Hessian uncertainties separately.
The comparisons are shown for all three energies of the BESIII measurements and for various bins in Bjorken-$x$ and inelasticity $y$ of the COMPASS measurements.
The theoretical calculations are systematically below the measurements of BESIII, especially for the pion production at 3.05~GeV.
However, the deficits can be largely compensated by the shifts due to both the correlated experimental uncertainties and the scale variations.
On the other hand, the NNLO calculations mostly overshoot the measurements of COMPASS with scale variations to be comparable in size.
Note for both the BESIII and COMPASS data the experimental uncertainties are dominated by correlated systematic uncertainties.
The results are similar for $\pi^-$ and $K^-$ production. 
In conclusion, we find that QCD collinear factorization can simultaneously describe both SIA and SIDIS processes at $Q$ values of a few GeV.

\begin{figure}[t]
    \centering
    \includegraphics[width=1.0\linewidth]{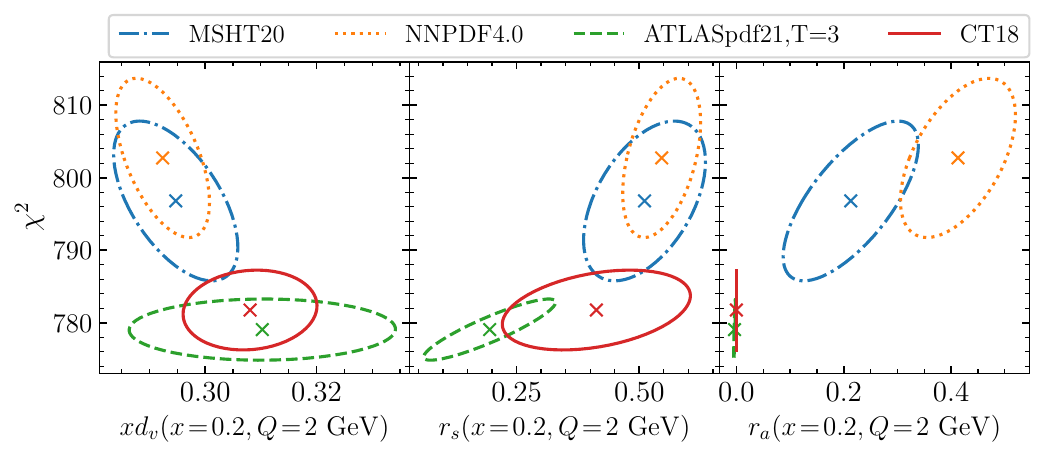}
    \caption{Correlation between the $\chi^2$ of fit to the hadron production data and PDF values $d_v, r_s, r_a$ (see text for details).}
    \label{fig:pdf}
\end{figure}

\noindent{\it Constraints on proton PDFs.--}
We further explore complementary constraints to the state-of-the-art PDFs at NNLO, especially focusing on constraints to PDFs of the strange quarks which are less known and have been actively studied recently~\cite{Gao:2017yyd,Amoroso:2022eow,Hou:2022onq}. 
{We have limited to an impact study using profile of log-likelihood functions rather than carrying out a full simultaneous fit of FFs and PDFs.
The latter can be done similarly as in Ref.~\cite{Moffat:2021dji} (at NLO accuracy) but requires far more efforts.}
Specifically, for the production of charged kaons in SIDIS, the differential cross sections receive large contributions from strange-quark PDFs due to the enhancement from fragmentation functions of strange quarks comparing to $u$ and $d$ quarks as shown in Fig.~\ref{fig:ffs}.  
For instance, considering the difference of differential cross sections of $K^+$ and $K^-$ production at COMPASS, at LO it can be expressed as
\begin{align}
{\dd^3\sigma^{K^+-K^-}\over \dd x\dd y\dd z}\propto \, & 2(u_v(x)+d_v(x))(D^{K^+}_u(z)-D^{K^+}_{\bar u}(z))\nonumber \\
&+s_v(x)(D^{K^+}_s(z)-D^{K^+}_{\bar s}(z)),
\end{align}
where we have assumed iso-scalar target without nuclear corrections, and $u_v$, $d_v$ being PDFs of valence quarks and $s_v\equiv s-\bar s$ being the asymmetry of strange (anti-)quark PDFs in the proton.
The kinematic variables are Bjorken-$x$, inelasticity $y$ and hadron energy fraction $z$.
The COMPASS data thus are sensitive to the strange quark asymmetries in the proton giving the dominance of $D^{K^+}_{\bar s}$ over FFs from all other quarks.
We select three representative values of PDFs for illustrations, which are
\begin{equation}
d_v\equiv d-\bar d,\,\,\,\,r_s\equiv {s+\bar s\over \bar u+\bar d},\,\,\,\,r_a\equiv {s-\bar s\over s+\bar s},
\end{equation}
at a momentum fraction $x=0.2$ and a factorization scale $Q=2$~GeV.
{Full results on PDFs with dependence on $x$ are also available (see Supplemental Material~\cite{fPDF}). }
The central values and Hessian uncertainties are evaluated using various NNLO PDF sets, including CT18~\cite{Hou:2019efy}, MSHT20~\cite{Bailey:2020ooq}, NNPDF4.0~\cite{NNPDF:2021njg} and ATLASpdf21~\cite{ATLAS:2021vod}.
We repeat the prescribed fit of FFs using all above PDFs including their central PDF sets as well as error sets, and obtain the profiled $\chi^2$ (minimized wrt. FFs) as functions of PDFs.
Correlations between the profiled $\chi^2$ and each of the PDF variables are shown in Fig.~\ref{fig:pdf} with the error ellipses at 68\% confidence level.
Note that both the CT18 and the ATLAS21 PDFs assume $s=\bar s$ at the initial scale and there are only small asymmetry due to QCD evolutions~\cite{Hou:2019efy}.

We find the $\chi^2$ exhibits strong correlations with $r_s$ and $r_a$, and anti-correlations with $d_v$ for NNPDF4.0 and MSHT20 sets.
We further take them as baseline PDFs for impact study using PDF reweighting for NNPDF4.0 and PDF profiling for MSHT20~\cite{Gao:2017yyd}, respectively.
In the case of NNPDF4.0 a standard Gaussian weight ($e^{-\chi^2/2}$) is used~\cite{Gao:2017yyd}, while a tolerance of $\Delta\chi^2=10$ is used in the profiling as suggested by the MSHT20 group~\cite{Bailey:2020ooq}.
The results are summarized in Tab.~\ref{tab:pdf} for the three PDF values from the original PDFs and after the reweighting or profiling.
The impact is more pronounced for $r_s$ and $r_a$ than for $d_v$ PDFs.
The NNPDF4.0 predicts a strange (anti-)quark asymmetry of more than 40\% at $x=0.2$, deviating from 0 with 3.8$\sigma$ significance, while the asymmetry is 21\% for MSHT20. 
Inclusion of the COMPASS SIDIS data reduces the asymmetry to about 28\% for NNPDF4.0, and to 17\% for MSHT20.
The sea-quark ratio $r_s$ is also reduced in both cases which is in agreement with preference from data on dimuon production~\cite{Liu:2022plj,Gao:2017kkx,Berger:2016inr}. 
The changes of PDF uncertainties are moderate (mild) for the case of NNPDF4.0 (MSHT20).

\begin{table*}
  \begin{tabular}{|c|c|c|c|}
    \hline
    & $xd_v (x = 0.2, Q = 2 \textrm{\rm{GeV}})$ & $r_s (x = 0.2, Q = 2
    \textrm{\rm{GeV}})$ & $r_a (x = 0.2, Q = 2 \textrm{\rm{GeV}})$\\
    \hline
    NNPDF4.0 & $0.2924 \pm 0.0084$ \  & $0.547 \pm 0.079$ & $0.408 \pm 0.107$\\
    \hline
    NNPDF4.0(reweighting) & $0.3021 \pm 0.0069$  & $0.438 \pm 0.066$  &
    $0.281 \pm 0.086$\\
    \hline
    MSHT20 & $0.295 \pm 0.011$ & $0.511 \pm 0.124$ & $0.213 \pm 0.126$ \\
    \hline
    MSHT20(profiling) & $0.298 \pm 0.011$ & $0.481 \pm 0.121$ & $0.167 \pm
    0.136$ \\
    \hline
  \end{tabular}
  \caption{PDF values $d_v, r_s, r_a$ at 68\% C.L. before and after reweighting
  (profiling) for NNPDF4.0 (MSHT20) PDF sets. 
  }
  \label{tab:pdf}
\end{table*}

\noindent{\it Conclusions.--}
We present a comprehensive global analysis of the fragmentation functions for identified charged hadrons at full NNLO and find good agreement with hadron multiplicities measured at low-$Q$ scales, around a few GeV, from both SIDIS and SIA processes.
{Our results indicate that QCD collinear factorization works well when hadrons are produced with energies greater than about $0.8$ ${\rm GeV}$, below which the fit quality deteriorate significantly.}
{The FFs are well-constrained for the favored quarks with world data.}
{Additionally, we investigate the sensitivity of SIDIS data to proton PDFs at NNLO accuracy, and find a preference for a reduced asymmetry in the strange (anti-)quark PDFs.}
Our work paves the way for future studies of nucleon structure by fully exploiting the data available from electron-ion colliders.

\quad \\
\noindent \textbf{Acknowledgments.} 
We would like to thank M. Stolarski for communications on the COMPASS data, G. Salam for communications on the HOPPET program, and L. Harland-Lang for providing us with the alternative MSHT20 PDF sets used for profiling.
The work of J.G. is supported by the National Natural Science Foundation of China (NSFC) under Grant No.~12275173 and open fund of Key Laboratory of Atomic and Subatomic Structure and Quantum Control (Ministry of Education). H. X. is supported by the NSFC under Grant No. 12475139. Y. Z. is supported by the NSFC under Grant No. U2032105 and CAS Project for Young Scientists in Basic Research Grant No. YSBR-117. X. S. is supported by the Helmholtz-OCPC Postdoctoral Exchange Program under Grant No. ZD2022004.

\noindent{\it Data availability.--} The data that support the findings of this article are openly available~\cite{lhapdf}.

% \bibliography{reference.bib}

\newpage
\widetext
\clearpage
\begin{center}
 \textbf{Supplemental Material for ``Fragmentation functions of charged hadrons at next-to-next-to-leading order and constraints on proton PDFs''}
\end{center} 

\subsection{Parameterization of the FFs}

In this work, the FFs at the starting scale $Q_0$ are parameterized as
\begin{equation}
z D_i^h \pqty{z, Q_0}
=
z^{\alpha_i^h} \pqty{1-z}^{\beta_i^h}
\exp(\sum_{n=0}^m a^h_{i,n} z^{n/2})
,
\end{equation}
where $i$ and $h$ label parton flavors and hadron species, respectively.
$\Bqty{\alpha_i^h, \beta_i^h, a^h_{i,n}}$ are the fitted parameters.
We increase $m$ until no discernible improvement in fit quality can be obtained.

The free parameters for $\pi^+$ and $K^+$ FFs are summarized in Table~\ref{tab:param-pion} and Table~\ref{tab:param-kaon}, respectively.
Certain flavor symmetries are assumed at the starting scale separately for favored quarks and un-favored quarks.
In addition, $\alpha$ and $\beta$ parameters, responsible for asymptotic behaviors of FFs, are assumed to be correlated for some flavors.  
That leads to a total number of 28 and 26 free parameters (indicated by check marks in the tables) for $\pi^+$ and $K^+$ FFs, respectively.

\begin{table}[h]  
\begin{tabular}{|c|c|c|c|c|c|c|}
\hline
flavor & favored & $a_0$ & $\alpha$ & $\beta$ & $a_1$ & $a_2$\\
\hline
$u=\overline{d}$ & \ding{51} & \ding{51} & \ding{51}  & \ding{51} & \ding{51} & \ding{51}\\
\hline
$d = \overline{u}$ & \ding{55} & \ding{51} & \ding{51}  & \ding{51} & \ding{51} & \ding{51}\\
\hline
$s = \overline{s}$ & \ding{55} & \ding{51} & $=\alpha_d$  & \ding{51} & \ding{51} & \ding{51}\\
\hline
$c = \overline{c}$ & \ding{55} & \ding{51} & \ding{51}  & \ding{51} & \ding{51} & \ding{51}\\
\hline
$b = \overline{b}$ & \ding{55} & \ding{51} & \ding{51}  & \ding{51} & \ding{51} & \ding{51}\\
\hline
$g$                & \ding{55} & \ding{51} & \ding{51}  & \ding{51} & \ding{51} & \ding{55}\\
\hline
\end{tabular}
  \caption{Parameters for the parton-to-$\pi^+$ FFs. 
  }
  \label{tab:param-pion}
\end{table}

\begin{table}[h]  
  \begin{tabular}{|c|c|c|c|c|c|c|}
\hline
flavor & favored & $a_0$ & $\alpha$ & $\beta$ & $a_1$ & $a_2$\\
\hline
$u$                & \ding{51} & \ding{51} & \ding{51}  & \ding{51} & \ding{51} & \ding{51}\\
\hline
$\overline{s}$     & \ding{51} & \ding{51} & $=\alpha_u$& $=\beta_u$ & \ding{51} & \ding{51}\\
\hline
$s = \overline{u}=d=\overline{d}$ & \ding{55} & \ding{51} & \ding{51}  & \ding{51} & \ding{51} & \ding{55}\\
\hline
$c = \overline{c}$ & \ding{55} & \ding{51} & \ding{51}  & \ding{51} & \ding{51} & \ding{51}\\
\hline
$b = \overline{b}$ & \ding{55} & \ding{51} & \ding{51}  & \ding{51} & \ding{51} & \ding{51}\\
\hline
$g$                & \ding{55} & \ding{51} & \ding{51}  & \ding{51} & \ding{51} & \ding{55}\\
\hline
\end{tabular}
  \caption{Similar to Table~\ref{tab:param-pion}, but for the parton-to-$K^+$ FFs. 
  }
  \label{tab:param-kaon}
\end{table}

%%%%%%%%%%%%%%%%%%%%%%%%%%%%%%%%%%%%%%%%%%%%%%%%%%%%%%%%%%%%%%%%%%%%%%%%%%%%%

%%%%%%%%%%%%%%%%%%%%%%%%%%%%%%%%%%%%%%%%%%%%%%%%%%%%%%%%%%%%%%%%%%%%%%%%%%%%%
\subsection{alternative fits} %pion-only and kaon-only analyses}
%
%%%%%%%%%%%%%%%%%%%%%%%%%%%%%%%%%%%%%%%%%%%%%%%%%%%%%%%%%%%%%%%%%%%%%%%%%%%%%
In this section, we explore various alternative fits.
To access the impact of fixed-order calculation accuracy, we present the fit quality of parallel fits at NLO in QCD in Table~\ref{tab:NLO-fit}.
Theoretical uncertainties are also estimated at NLO, which are in general larger than their NNLO counterparts.
The full NNLO predictions describe the data slightly better than the NLO ones 
when $E_{h, {\rm min}}\ge 0.7~{\rm GeV}$ case, while NNLO predictions slightly deteriorate the fit quality when $E_{h, {\rm min}}\le 0.6~{\rm GeV}$.
The FFs extracted at NLO accuracy according to our nominal choice
$E_{h, {\rm min}}=0.8~{\rm GeV}$ is presented in Fig.1 of the paper.

\begin{table}[h]
  \begin{tabular}{|c|c|c|c|c|c|c|c|c||c|c|c|}
    \hline
     \multirow{2}{*}{${E_{h,{\rm min}}}$$[{\rm{GeV}}]$} &  \multicolumn{2}{c|}{BESIII} &  \multicolumn{2}{c|} {COMPASS} & \multicolumn{2}{c|} {B-factories} & \multicolumn{2}{c||} {HE-SIA} &  \multicolumn{3}{c|} {global} \\
    \cline{2-12}
      & $N_{\rm{pt}}$ &
    ${\chi^2}/{N_{\rm{pt}}}$ & $N_{\rm{pt}}$ &
    ${\chi^2}/{N_{\rm{pt}}}$ & $N_{\rm{pt}}$ &
    ${\chi^2}/{N_{\rm{pt}}}$ & $N_{\rm{pt}}$ & 
    ${\chi^2}/{N_{\rm{pt}}}$ & $N_{\rm{pt}}$ & $\chi^2$ &
    ${\chi^2}/{N_{\rm{pt}}}$\\
    \hline  
    0.5 & 242 & 1.38 & 358 & 1.50 & 233 & 1.01 & 426 & 1.23 & 1259 & 1631.2 &
    1.30\\
    0.6 & 212 & 1.26 & 290 & 1.44 & 228 & 0.87 & 423 & 1.06 & 1153 & 1333.2 &
    1.16\\
    0.7 & 182 & 1.12 & 214 & 1.43 & 223 & 0.67 & 413 & 0.97 & 1032 & 1057.9 &
    1.03\\
    0.8 & 152 & 1.03 & 142 & 1.26 & 218 & 0.54 & 407 & 0.85 & 919 & 801.6 &
    0.87\\
    0.9 & 122 & 1.08 & 94 & 1.22 & 213 & 0.52 & 407 & 0.84 & 836 & 697.5 &
    0.83\\
    1.0 & 98 & 1.18 & 54 & 0.93 & 209 & 0.49 & 403 & 0.83 & 764 & 603.7 &
    0.79\\
    \hline
  \end{tabular}
  \caption{Fit quality for NLO fit with different choices of lower cut on the hadron energy.}
  \label{tab:NLO-fit}  
\end{table}

%%%%%%%%%%%%%%%%%%%%%%%%%%%%%%%%%%%%%%%%%%%%%%%%%%%%%%%%%%%%%%%%%%%%%%%%%%%%%
In Table~\ref{tab:fit-pion} and Table~\ref{tab:fit-kaon}, we present the fit quality of FFs determination using either pion or kaon data, 
in contrast to our nominal fit which determines pion and kaon FFs simultaneously.
For both the pion-only and kaon-only fits,
quick growth of $\chi^2/N_{\rm pt}$ is observed when $E_{h, {\rm  min}} < 0.8~{\rm GeV}$.
We also find better overall agreements and faster average growth of $\chi^2/N_{\rm pt}$ for kaons than for pions.

\begin{table}[h]
  \begin{tabular}{|c|c|c|c|c|c|c|c|c||c|c|c|}
    \hline
     \multirow{2}{*}{${E_{h,{\rm min}}}$$[{\rm{GeV}}]$} &  \multicolumn{2}{c|}{BESIII} &  \multicolumn{2}{c|} {COMPASS} & \multicolumn{2}{c|} {B-factories} & \multicolumn{2}{c||} {HE-SIA} &  \multicolumn{3}{c|} {global} \\
    \cline{2-12}
      & $N_{\rm{pt}}$ &
    ${\chi^2}/{N_{\rm{pt}}}$ & $N_{\rm{pt}}$ &
    ${\chi^2}/{N_{\rm{pt}}}$ & $N_{\rm{pt}}$ &
    ${\chi^2}/{N_{\rm{pt}}}$ & $N_{\rm{pt}}$ & 
    ${\chi^2}/{N_{\rm{pt}}}$ & $N_{\rm{pt}}$ & $\chi^2$ &
    ${\chi^2}/{N_{\rm{pt}}}$\\
    \hline
    0.5 & 112 & 1.68 & 180 & 1.70 & 113 & 0.81 & 230 & 1.10 & 635 & 841.5 &
    1.33\\
    0.6 & 100 & 1.59 & 146 & 1.48 & 111 & 0.63 & 227 & 0.97 & 584 & 667.7 &
    1.14\\
    0.7 & 88 & 1.31 & 108 & 1.39 & 109 & 0.63 & 222 & 0.90 & 527 & 535.5 &
    1.02\\
    0.8 & 76 & 1.09 & 72 & 1.23 & 107 & 0.61 & 219 & 0.85 & 474 & 424.2 &
    0.89\\
    0.9 & 64 & 1.11 & 48 & 1.37 & 105 & 0.60 & 219 & 0.84 & 436 & 383.6 &
    0.88\\
    1.0 & 52 & 1.10 & 28 & 1.24 & 103 & 0.56 & 217 & 0.84 & 400 & 331.6 &
    0.83\\
    \hline
  \end{tabular}
  \caption{Fit quality of pion-only analyses for different choices of lower cut on the hadron energy.}
  \label{tab:fit-pion}
\end{table}

\begin{table}[h]
  \begin{tabular}{|c|c|c|c|c|c|c|c|c||c|c|c|}
    \hline
     \multirow{2}{*}{${E_{h,{\rm min}}}$$[{\rm{GeV}}]$} &  \multicolumn{2}{c|}{BESIII} &  \multicolumn{2}{c|} {COMPASS} & \multicolumn{2}{c|} {B-factories} & \multicolumn{2}{c||} {HE-SIA} &  \multicolumn{3}{c|} {global} \\
    \cline{2-12}
      & $N_{\rm{pt}}$ &
    ${\chi^2}/{N_{\rm{pt}}}$ & $N_{\rm{pt}}$ &
    ${\chi^2}/{N_{\rm{pt}}}$ & $N_{\rm{pt}}$ &
    ${\chi^2}/{N_{\rm{pt}}}$ & $N_{\rm{pt}}$ & 
    ${\chi^2}/{N_{\rm{pt}}}$ & $N_{\rm{pt}}$ & $\chi^2$ &
    ${\chi^2}/{N_{\rm{pt}}}$\\
    \hline  
    0.5 & 130 & 1.03 & 178 & 1.58 & 120 & 1.39 & 196 & 1.26 & 624 & 828.5 &
    1.33\\
    0.6 & 112 & 0.97 & 144 & 1.54 & 117 & 1.23 & 196 & 0.99 & 569 & 669.7 &
    1.18\\
    0.7 & 94 & 0.95 & 106 & 1.46 & 114 & 0.66 & 191 & 0.76 & 505 & 464.4 &
    0.92\\
    0.8 & 76 & 0.97 & 70 & 1.36 & 111 & 0.47 & 188 & 0.74 & 445 & 360.2 &
    0.81\\
    0.9 & 58 & 1.02 & 46 & 1.23 & 108 & 0.46 & 188 & 0.73 & 400 & 303.1 &
    0.76\\
    1.0 & 46 & 1.20 & 26 & 0.86 & 106 & 0.44 & 186 & 0.72 & 364 & 258.4 &
    0.71\\
    \hline
  \end{tabular}
  \caption{Similar to Table~\ref{tab:fit-pion} but for kaon-only analyses.} 
  \label{tab:fit-kaon}
\end{table}

%%%%%%%%%%%%%%%%%%%%%%%%%%%%%%%%%%%%%%%%%%%%%%%%%%%%%%%%%%%%%%%%%%%%%%%%%%%%%

In order to study the impact of BESIII data sets, we also performed alternative fits with BESIII data sets excluded.
We summarize the fit quality in Table~\ref{tab:fit-noBES}. 
The description of
COMPASS data is slightly improved when excluding the
BESIII data from the fit.
The resulting FFs according to $E_{h,{\rm min}} = 0.8~{\rm GeV}$ have been presented in Fig.~1 of the paper.

\begin{table*}[h]
  \begin{tabular}{|c|c|c|c|c|c|c||c|c|c|}
    \hline
     \multirow{2}{*}{${E_{h,{\rm min}}}{[\rm{GeV}}]$} &  \multicolumn{2}{c|} {COMPASS} & \multicolumn{2}{c|} {B-factories} & \multicolumn{2}{c||} {HE-SIA} &  \multicolumn{3}{c|} {global} \\
    \cline{2-10}
      & $N_{\rm{pt}}$ &
    ${\chi^2}/{N_{\rm{pt}}}$ & $N_{\rm{pt}}$ &
    ${\chi^2}/{N_{\rm{pt}}}$ & $N_{\rm{pt}}$ & 
    ${\chi^2}/{N_{\rm{pt}}}$ & $N_{\rm{pt}}$ & $\chi^2$ &
    ${\chi^2}/{N_{\rm{pt}}}$\\
    \hline  
    0.5 & 358 & 1.49 & 233 & 0.97 & 426 & 1.03 & 1017 & 1200.0 & 1.18\\
    0.6 & 290 & 1.37 & 228 & 0.82 & 423 & 0.94 & 941 & 983.1 & 1.04\\
    0.7 & 214 & 1.34 & 223 & 0.61 & 413 & 0.84 & 850 & 771.2 & 0.91\\
    0.8 & 142 & 1.23 & 218 & 0.52 & 407 & 0.81 & 767 & 620.2 & 0.81\\
    0.9 & 94 & 1.22 & 213 & 0.50 & 407 & 0.80 & 714 & 549.0 & 0.77\\
    1.0 & 54 & 0.95 & 209 & 0.47 & 403 & 0.80 & 666 & 472.0 & 0.71\\
    \hline
  \end{tabular}
  \caption{Fit quality of NNLO analyses with all BESIII data sets excluded. 
  }
  \label{tab:fit-noBES}
\end{table*}

%%%%%%%%%%%%%%%%%%%%%%%%%%%%%%%%%%%%%%%%%%%%%%%%%%%%%%%%%%%%%%%%%%%%%%%%%%%%%
\subsection{Comparison to other groups}

\begin{figure}[hbp]
  \centering
  \includegraphics[width=0.8\textwidth]{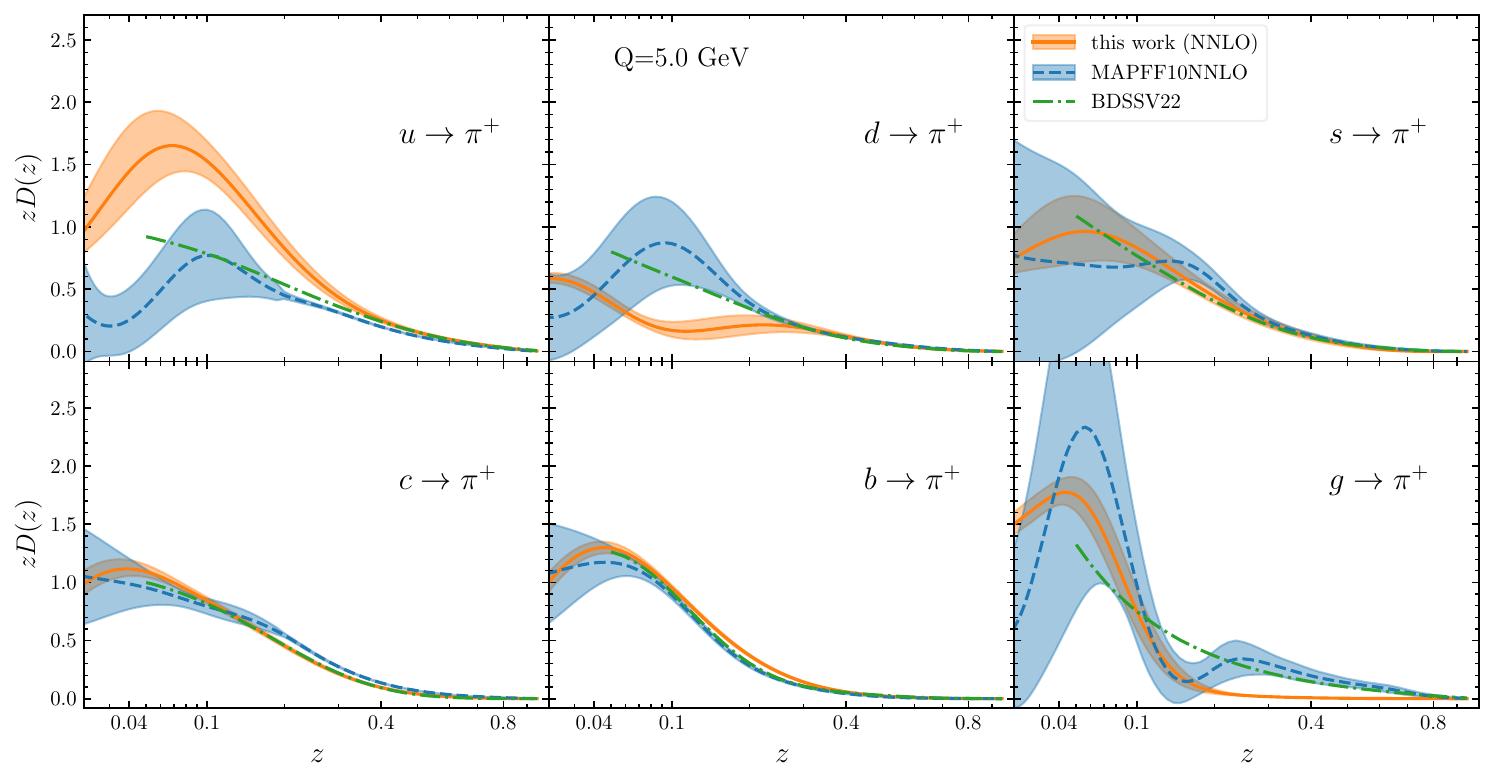}
	\caption{
        Comparison of the FFs obtained in this work with those from BDSSV22~\cite{Borsa:2022vvp} and MAPFF10~\cite{AbdulKhalek:2022laj}
        at $Q=5.0~{\rm GeV}$.
	}
  \label{fig:SM-FFs}
\end{figure}

We compare FFs obtained in this work with FFs from BDSSV22~\cite{Borsa:2022vvp} and MAPFF10~\cite{AbdulKhalek:2022laj} at $Q=5.0~{\rm GeV}$ in Fig.~\ref{fig:SM-FFs}.
Both BDSSV22 and MAPFF10 FFs are determined at approximate NNLO accuracy, with the former providing only pion FFs at $z>0.05$.
Good agreements are found for the $u/d$ quark FFs at large-$z$ and the $s$ quark FFs.
The gluon FFs are not well constrained due to the absence of $pp$-collision data in all these analyses.

%%%%%%%%%%%%%%%%%%%%%%%%%%%%%%%%%%%%%%%%%%%%%%%%%%%%%%%%%%%%%%%%%%%%%%%%%%%%%
\subsection{Impact of SIDIS data on the proton PDFs}%The modified PDF sets}

\begin{figure}[hbp]
  \centering
  \includegraphics[width=0.8\textwidth]{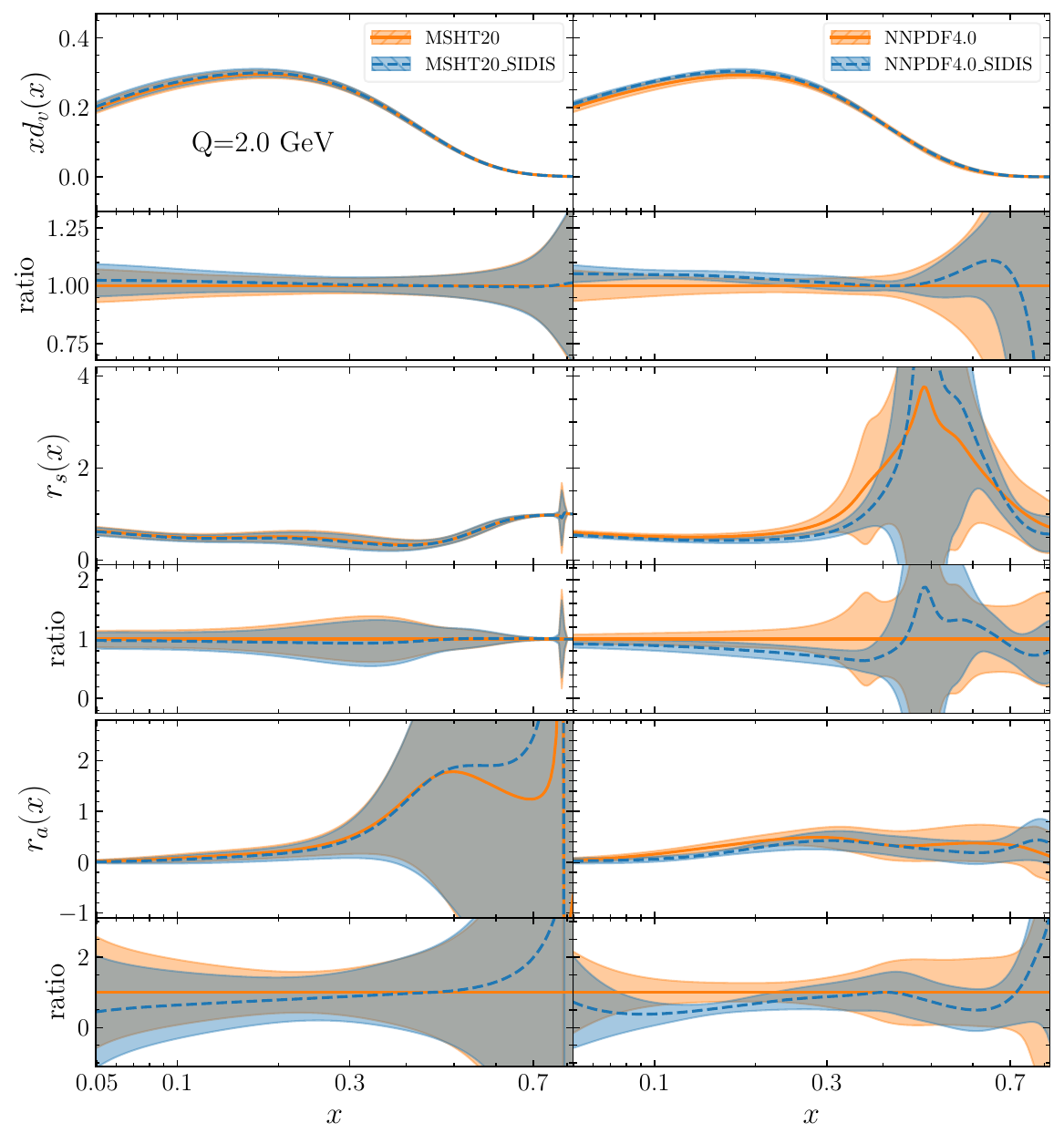}
	\caption{
        Comparison of the original MSHT20 and NNPDF4.0 PDF sets with the ones after profiling or reweighting with SIDIS data.
	}
  \label{fig:SM-PDFs}
\end{figure}

In order to study the impact of SIDIS data on proton PDFs using baseline of MSHT20~\cite{Bailey:2020ooq} (with tolerance $T^2=10$) and NNPDF4.0~\cite{NNPDF:2021njg} PDF sets, 
we compare the 
PDFs after profling/reweighting with the original ones in Fig.~\ref{fig:SM-PDFs} at $Q=2.0~{\rm GeV}$ as functions of $x$.
The PDF values $d_v(x), r_s(x), r_a(x)$ are defined in Eq.(2) of the paper.
As we can see, SIDIS data prefers a small $r_a$ value for moderate and low $x$.
Note that the COMPASS SIDIS data covers kinematic regions with Bjorken-$x$ variable, $0.05<x_B<0.4$.

\end{document}